%% file: main.tex
\documentclass[sigconf]{acmart}
\usepackage{cleveref}

\usepackage{bm}
\usepackage{bbm}
\usepackage{caption}
\usepackage{subcaption}
\usepackage{booktabs,amsfonts,dcolumn}
\usepackage{mathtools}
\usepackage{wasysym}
\usepackage{hhline}
\usepackage{float}
\usepackage{commath}

 \usepackage{multirow}
 \usepackage[ruled,vlined,linesnumbered]{algorithm2e}
\usepackage{booktabs,siunitx}

\usepackage{stmaryrd}
\usepackage{trimclip}
\usepackage{pifont}
\usepackage{enumitem}
\usepackage{mleftright}
\usepackage{scalerel,stackengine}
\usepackage{bbold}

\renewcommand\footnotetextcopyrightpermission[1]{} 
\setcopyright{acmlicensed}
\copyrightyear{2018}
\acmYear{2018}
\acmDOI{XXXXXXX.XXXXXXX}
\acmConference[Conference acronym 'XX]{Make sure to enter the correct
  conference title from your rights confirmation email}{June 03--05,
  2018}{Woodstock, NY}
\acmISBN{978-1-4503-XXXX-X/18/06}

\begin{document}
\title{Structure-aware Relative Policy Optimization for Ranking}

\author{Yiteng Tu}
\affiliation{%
    \institution{Tsinghua University}
    \city{Beijing}
    \country{China}
}
\email{tyt24@mails.tsinghua.edu.cn}

\author{Weihang Su}
\affiliation{%
    \institution{Tsinghua University}
    \city{Beijing}
    \country{China}
}

\author{Zitao Su}
\affiliation{%
    \institution{Renmin University of China}
    \city{Beijing}
    \country{China}
}

\author{Yiqun Liu}
\affiliation{%
    \institution{Tsinghua University}
    \city{Beijing}
    \country{China}
}

\author{Min Zhang}
\affiliation{%
    \institution{Tsinghua University}
    \city{Beijing}
    \country{China}
}

\author{Qingyao Ai}
\authornote{Corresponding Author}
\affiliation{%
    \institution{Tsinghua University}
    \city{Beijing}
    \country{China}
}

\input{secs/0abs}

\ccsdesc[500]{Information systems~Learning to rank}
\ccsdesc[500]{Information systems~Retrieval models and ranking}
\keywords{Ranking, Reinforcement Learning, Structure-aware}

\maketitle

\input{secs/1intro}
\input{secs/2related}
\input{secs/3method}

\input{secs/4experi}

\input{secs/5result}
\input{secs/6conclu}

\bibliographystyle{ACM-Reference-Format}
\bibliography{sample-base}

\input{secs/7appendix}

\end{document}

%% file: secs/0abs.tex
\begin{abstract}
Ranking is a fundamental component of modern information access systems. 
Reinforcement learning (RL) provides a flexible framework for directly optimizing coarse-grained feedback and system-level objectives defined over the complete ranking list. 
However, existing RL-based ranking methods typically treat each sampled permutation as an atomic output and evaluate it primarily through a scalar reward, overlooking the structural relationships among different ranking lists. 
Consequently, permutations with similar rewards but substantially different permutation patterns may receive comparable optimization signals, potentially leading to inaccurate credit assignment and overly aggressive policy updates. 
To address this limitation, we propose SRPO, a \textbf{S}tructure-aware \textbf{R}elative \textbf{P}olicy \textbf{O}ptimization framework for listwise ranking. 
SRPO measures the discrepancy between sampled permutations using a top-weighted Kendall-tau distance and normalizes their pairwise reward differences by the corresponding distances. 
It quantifies the reward improvement per unit of ranking change, thereby emphasizing efficient local refinements, particularly those involving top-ranked positions. 
Experimental results across two ranking scenarios demonstrate that explicitly modeling permutation-level differences improves the effectiveness and stability of listwise ranking, with particularly favorable performance in limited-feedback and complex list-level optimization settings.
Our code is available at: \url{https://github.com/StibiumT16/SRPO-Rank}.
\end{abstract}

%% file: secs/1intro.tex
\section{Introduction}
Ranking is a fundamental component of modern information access systems, including web search~\cite{liu2025e2rank,ai2018learning,guo2020deep}, recommendation~\cite{chen2019top,zhang2017joint,chen2017personalized}, and online advertising~\cite{gu2026deep,feng2007ranked,karimzadehgan2011stochastic}. 
However, in real-world scenarios, most existing ranking methods are trained with fine-grained candidate-level relevance judgments and optimize pointwise, pairwise, or listwise surrogate objectives based on such labels~\cite{liu2009learning,cao2007learning,xia2008listwise,burges2006learning,burges2005learning,burges2010ranknet}. 
Although these approaches have achieved strong empirical performance, obtaining high-quality relevance annotations is costly and difficult to scale~\cite{snow2008cheap,chapelle2009expected}.
Moreover, such fine-grained relevance labels do not always directly reflect the ultimate objectives of real-world ranking systems, which may involve user clicks, engagement, diversity, provider fairness, or other system-level considerations.
In contrast, coarse-grained feedback collected at the ranking list, session, or search engine result page (SERP) level, such as dwell time, query reformulation, session engagement, and exposure fairness, is often easier to obtain and more closely aligned with practical ranking objectives~\cite{dupret2013absence,liu2014skimming,joachims2007evaluating,ai2018unbiased,joachims2017accurately,ai2021unbiased}.

However, learning a ranking policy from such coarse-grained feedback is challenging because signals defined over the entire ranking list can be non-differentiable and noisy, and they generally cannot be decomposed into explicit supervision for individual candidates~\cite{tu2022reinforcement}. 
Fortunately, reinforcement learning (RL) provides a natural framework for this setting by treating a ranking model as a stochastic policy over permutations and directly optimizing the expected listwise reward. 
Existing listwise RL-based ranking methods commonly employ a policy-gradient algorithm~\cite{sutton1999policy,silver2014deterministic} to sample ranking lists and update the policy based on observed rewards~\cite{singh2019policy,tu2022reinforcement,oosterhuis2018ranking}.
Nevertheless, classical policy-gradient methods often suffer from high-variance gradient estimation and limited sample efficiency~\cite{gu2016q,greensmith2004variance,schulman2015high}.
On the other hand, recent progress in RL for large language models (LLMs)~\cite{guo2025deepseek,hurst2024gpt,yang2025qwen3,achiam2023gpt,zheng2025group} has introduced more stable optimization algorithms based on relative comparisons among multiple sampled outputs. 
In particular, Group Relative Policy Optimization (GRPO)~\cite{guo2025deepseek,shao2024deepseekmath} estimates the advantage of each output by normalizing its reward against a group of samples generated for the same input, avoiding the need to train an additional value model. 
Prior work~\cite{tu2022reinforcement} has adapted this algorithm to listwise ranking by sampling multiple permutations for each query from a Plackett–Luce distribution~\cite{plackett1975analysis,luce1959individual}. 
The resulting approach demonstrates that modern RL algorithms can effectively optimize ranking models using coarse-grained listwise rewards and substantially outperform conventional policy-gradient methods.

\begin{figure}[t]
    \centering
    \includegraphics[width=0.5\textwidth]{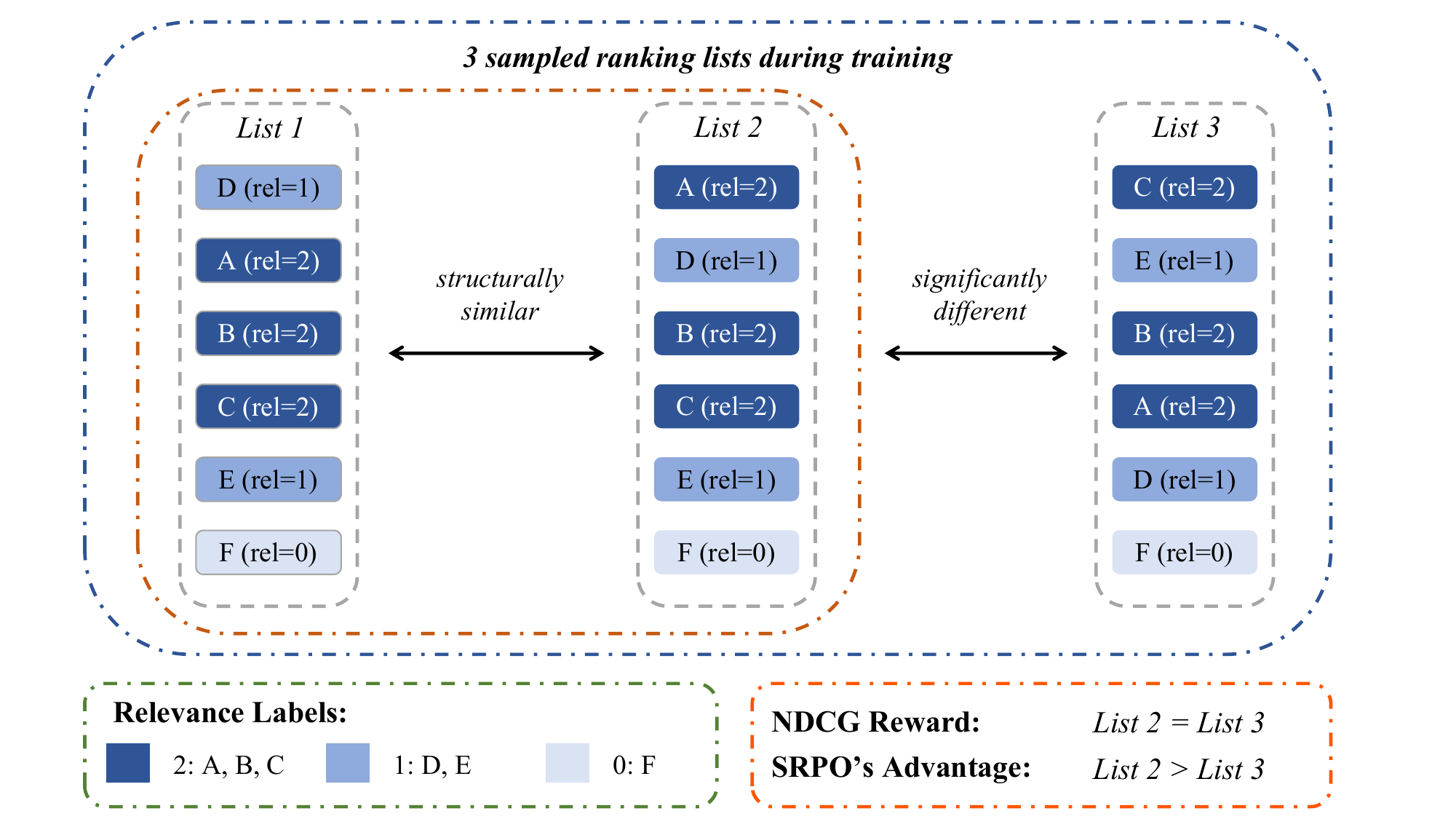}
    \caption{An example demonstrates the impact of the ranking structure on the advantage function in RL-based ranking.}
    \label{fig:main}
\end{figure}

Despite their effectiveness, all these classical and recent RL algorithms, including GRPO, do not explicitly account for the structural characteristics of ranking. 
When applied directly to ranking, they treat each sampled permutation as an atomic output and compare ranking lists solely based on their scalar rewards. 
However, ranking takes place in a highly structured permutation space~\cite{xia2008listwise,cao2007learning,korba2018structured}, where the contrast between two ranking lists depends not only on their rewards, such as ranking metrics, but also on the positions of their candidates. 
Two permutations with similar rewards may differ substantially in structure. 
In contrast, a small local swap, especially near the top of the list, may lead to a meaningful improvement in ranking quality.
For example, in Figure~\ref{fig:main}, List 1 (L1) and List 2 (L2) differ by only a single local swap at the top positions, and therefore share highly similar structural patterns, while List 3 (L3) exhibits substantially larger permutation-level deviation from the other two.
Although all three lists may achieve comparable listwise reward values (e.g., NDCG: L2 = L3 > L1), their structural properties are clearly different.
From a policy stability perspective, L2 and L3 should not be treated as equally important, since L1 and L2 share similar structures and are thus more compatible with the ranking score distribution.
In contrast, L3 achieves a similar reward by introducing a much more aggressive and disruptive reordering of the permutation; overemphasizing it may lead to instability in strategy training.
Therefore, reward differences alone cannot capture the structural cost of transforming one ranking into another, nor can they distinguish whether a performance gain is achieved through a meaningful local refinement, an arbitrary global permutation, or a locally superior anomaly sampling. 
Therefore, treating all reward improvements equally without considering the permutational structure may lead to overly aggressive updates when high-reward but structurally inconsistent rankings are sampled, thereby potentially destabilizing the optimization process and degrading the learned ranking policy.
In this sense, existing RL-based ranking methods are mostly reward-aware but structure-agnostic: they identify which ranking is better, but do not model how the ranking structure changes to achieve the improvement.

To address this challenge, we propose Structure-aware Relative Policy Optimization (SRPO), a policy optimization method for listwise RL-based ranking. 
The central idea of SRPO is to evaluate reward improvement jointly with the structural discrepancy between sampled permutations. 
For each pair of rankings in a sampled group, SRPO measures their discrepancy using a top-weighted Kendall-tau distance~\cite{kendall1938new} and normalizes their reward difference by this distance. 
Thus, the resulting structure-normalized preference represents the reward improvement achieved per unit of ranking modification and therefore emphasizes efficient local refinements, particularly those involving high-ranked positions. 
SRPO then aggregates pairwise preferences within each sampled group to construct a contrastive relative advantage, which naturally centers the advantages across the group.
Consequently, SRPO retains the candidate-level metric-agnostic nature of coarse-grained feedback while aligning the optimization process more closely with the sequential and combinatorial structure of ranking.
We evaluate SRPO in two scenarios: learning to rank (LTR) and LLM-based text reranking. 
Experimental results show that explicitly modeling permutation-level differences improves the effectiveness and stability of relative policy optimization for ranking. 
Specifically, SRPO achieves strong performance across different datasets and ranking objectives, while exhibiting favorable behavior under limited sampling and complex list-level rewards.

This work makes three main contributions. 
First, we identify an important limitation of existing RL-based ranking methods: their relative advantages depend only on scalar rewards and overlook the structural relationships among ranking permutations. 
Second, we propose SRPO, which integrates top-weighted permutation distance, structure-normalized pairwise preferences, and contrastive advantage estimation into a unified RL framework for listwise ranking. 
Third, extensive experiments demonstrate the effectiveness and generality of SRPO across varying settings and feedback regimes.

%% file: secs/2related.tex
\section{Related Work}
\subsection{Reinforcement Learning to Rank (RLTR)}
Reinforcement learning (RL) studies how an agent learns a policy through interaction with an environment to maximize expected cumulative rewards~\cite{sutton1998reinforcement,kaelbling1996reinforcement}. 
Reinforcement learning to rank (RLTR) formulates ranking as a sequential decision-making or stochastic policy optimization problem, allowing ranking models to directly optimize item-level feedback or non-differentiable evaluation objectives~\cite{tu2022reinforcement}. 
Early RLTR approaches commonly model the construction of a ranking list as a Markov decision process (MDP) in which documents are selected sequentially and candidate-level rewards are assigned to individual decisions~\cite{wei2017reinforcement,chen2019top,xu2020reinforcement}. 
However, these methods are still limited by candidate-level supervised signals.
Another line of work~\cite{singh2019policy,oosterhuis2021computationally,oosterhuis2022learning} models a ranking policy through the Plackett–Luce distribution~\cite{plackett1975analysis,luce1959individual} and directly optimizes expected ranking utility using sampled permutations. 
These methods have also been extended to more sophisticated ranking settings beyond relevance like exposure fairness~\cite{singh2019policy,morik2020controlling,yang2021maximizing,tu2026equity}, where RL provides a natural way to optimize such global exposure objectives that are difficult to express as conventional supervised losses. 
More recently, advanced RL algorithms like GRPO~\cite{shao2024deepseekmath,guo2025deepseek} have been adapted for listwise RLTR, which substantially improves optimization stability and ranking performance, demonstrating that ranking models can be effectively learned from coarse-grained listwise feedback and even outperform supervised learning methods based on costly candidate-level feedback~\cite{tu2022reinforcement}.
Nevertheless, all these methods compare sampled rankings only through scalar rewards and treat complete permutations as atomic actions, overlooking the structural relationships among rankings. 
In contrast, our SRPO explicitly measures permutation-level discrepancies and uses them to construct structure-aware relative advantages and finer-grained policy updates.

\subsection{LLM-based Text Reranker}
LLMs have recently been explored as powerful text rerankers due to their strong semantic understanding and instruction-following capabilities~\cite{ke2026resrank,liu2025e2rank,sun2023chatgpt,zhuang2025rank}.
Compared with traditional neural ranking models, LLM-based rerankers can better capture fine-grained query-document interactions and perform listwise comparison among multiple candidates~\cite{sun2023chatgpt}.
Existing studies can be broadly categorized into pointwise, pairwise, and listwise paradigms. 
Pointwise methods independently estimate the relevance of each query-document pair, typically by prompting or fine-tuning LLMs to generate relevance scores or judgments~\cite{nogueira2019multi,chen2024bge,nogueira2020document}. 
Pairwise methods instead ask LLMs to compare two candidate documents at a time, converting relative preferences into a final ranking through aggregation or sorting procedures~\cite{qin2024large}. 
Listwise methods further extend this idea by presenting multiple candidates simultaneously and prompting the model to directly produce an ordered list, thereby better matching the nature of ranking as a permutation prediction problem~\cite{sun2023chatgpt,pradeep2023rankzephyr,pradeep2023rankvicuna}. 
While these approaches have substantially advanced reranking, most of them rely on supervised labels, preference signals, or scalar list-level feedback, and rarely model the structural discrepancy among different output permutations.

%% file: secs/3method.tex
\begin{figure*}[t]
    \centering    
    \includegraphics[width=1.0\textwidth]{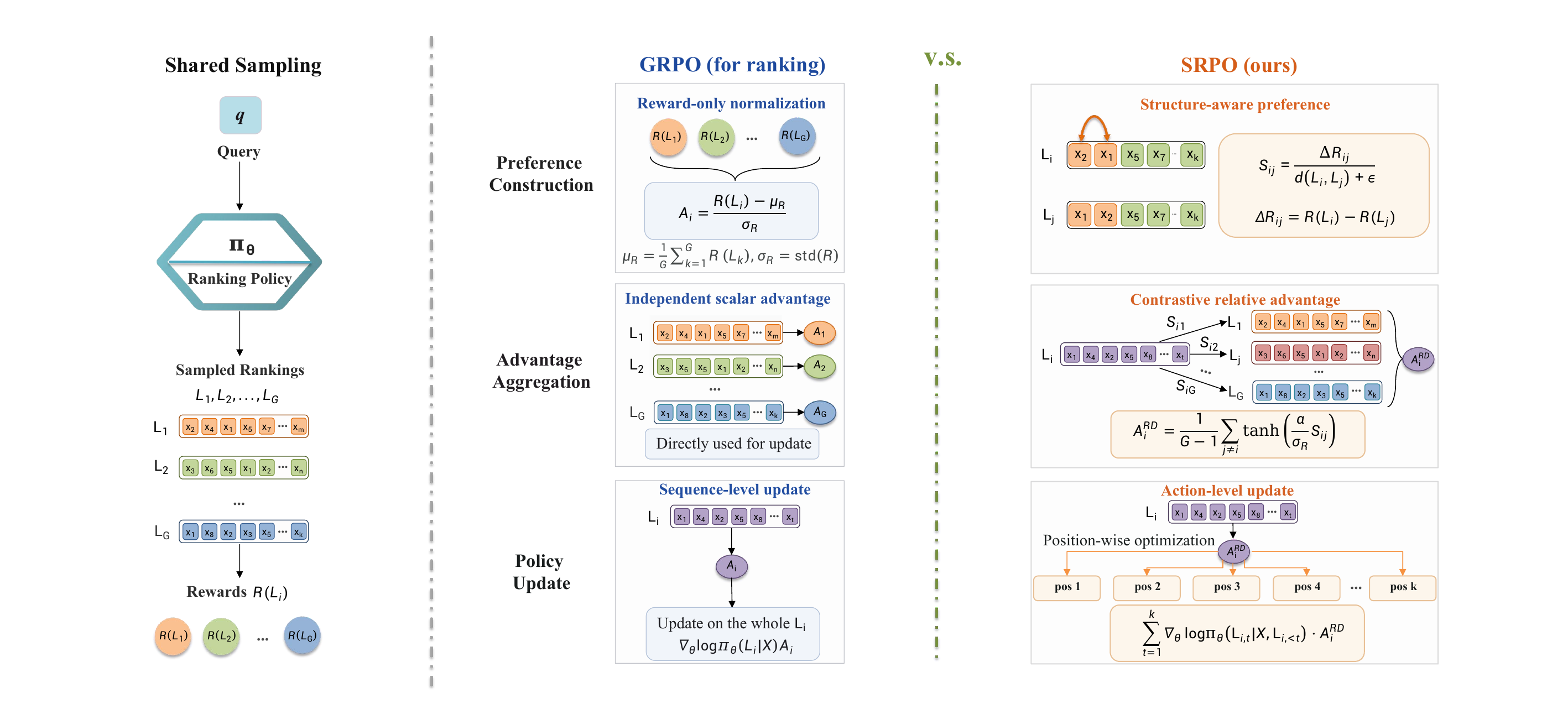}
    \vspace{-4mm}
    \caption{An illustration of our SRPO and its comparison with GRPO (tailored for ranking in \cite{tu2022reinforcement}).}    \label{fig:method}
\end{figure*}

\section{Methodology}
\subsection{Preliminaries}
\subsubsection{Listwise RL-based Ranking} \hfill\newline
Following recent studies~\cite{tu2022reinforcement,singh2019policy,gao2023policy}, we formulate ranking as a listwise policy optimization problem.
Given a candidate set $X=(x_1,\ldots,x_k)$ represented by query-document pairs for reranker or feature vectors for LTR, a ranking model $h_\theta$ assigns a relevance score to each candidate.
This score can either be output directly by a neural network or interpreted as the inner product of the query and document embeddings.
The objective is to learn a ranking policy $\pi_\theta$ that maximizes the expected listwise reward:
\begin{equation}
    \max_{\theta}\;
    \mathbb{E}_{L\sim \pi_\theta(\cdot|X)} \mathcal{R}(L),
\end{equation}
where $L$ denotes a sampled ranking list with size $k$ and $\mathcal{R}(L)$ is a listwise reward defined on the entire ranking.
To enable listwise policy optimization, we adopt the Plackett-Luce (PL) ranking model~\cite{plackett1975analysis,luce1959individual}, which defines a probability distribution over ranking permutations. 
Specifically, given the scoring function $h_\theta$, the score vector for a candidate set $X$ is $h_\theta(X)=\left(h_\theta(x_1),\ldots,h_\theta(x_k)\right)$.
Based on these scores, the probability of generating a ranking list
$L=(L[1],\ldots,L[k])$ under the PL policy is:
\begin{equation} \label{eq:pl}
\pi_{\theta}(L \mid X)
=
\prod_{i=1}^{k}
\frac{
\exp(h_\theta(x_{L[i]}))
}{
\sum_{j=i}^{k}
\exp(h_\theta(x_{L[j]}))
}.
\end{equation}
This formulation enables direct optimization using listwise reward signals via policy-gradient approaches~\cite{sutton1999policy,silver2014deterministic}.
Furthermore, with the help of the Gumbel Softmax trick~\cite{gumbel1954statistical}, we can achieve efficient sampling at $\mathcal{O}(k\log k)$ (see Appendix~\ref{app:pl} for details).

\subsubsection{Group Relative Policy Optimization (GRPO) for Ranking} \label{subsubsec:grpo} \hfill\newline
Recent advances in large language models (LLMs) have led to substantial progress in RL algorithms~\cite{shao2024deepseekmath,zheng2025group,yu2026dapo,schulman2017proximal,zheng2025group}.
Among them, Group Relative Policy Optimization (GRPO)~\cite{guo2025deepseek,shao2024deepseekmath} has emerged as an effective and stable policy optimization method based on group-wise relative comparisons.
\citet{tu2022reinforcement} first introduced GRPO into listwise ranking optimization with coarse-grained rewards, demonstrating strong performance under such list-level signals.
Given a candidate set $X$, GRPO samples a group of ranking permutations $\mathcal{G}=\{L_1,\ldots,L_G\}$ with group size $G$ from the current PL policy $\pi_\theta$.
The reward of each sampled permutation is normalized within the group to construct a relative advantage:
\begin{equation}
A_i = \frac{\mathcal{R}(L_i) - \text{mean}\left(\{\mathcal{R}(L_i)\}_{i=1}^G\right)}{\text{std}\left(\{\mathcal{R}(L_i)\}_{i=1}^G\right)}.
\end{equation}
The policy is then optimized using the group-relative objective (note that we adopt a simplified one-step formulation without clip operation following~\cite{tu2022reinforcement,zhuang2025rank}):
\begin{equation}
J_{\text{GRPO}}(\theta) = \mathbb{E}_{L_i \sim \pi_{\theta}} \frac{1}{G}\sum_{i=1}^G \left[r_{\theta}(L_i \mid X)  \cdot A_i - \beta \cdot \mathbb{D}_{\mathrm{KL}}(\pi_\theta || \pi_{\mathrm{ref}})\right], 
\end{equation}
where $r_{\theta}(L_i \mid X) = \frac{\pi_{\theta}(L_i\ \mid X)}{\pi_{\theta_{\text{old}}}(L_i\ \mid X)}$ is the importance weight, $\beta$ is the coefficient of the KL penalty $\mathbb{D}_{KL}(\pi_\theta || \pi_{ref})$, which penalizes the deviation between the new policy and the frozen reference policy.

Although GRPO effectively reduces gradient variance through group-wise normalization, it compares sampled rankings solely based on scalar reward values.
However, as discussed above, ranking operates in a highly structured permutation space, where scalar reward differences alone are insufficient to characterize the contrast between ranking lists, since permutations with comparable rewards may correspond to drastically different structural transformations.

\subsection{Our method: Structure-aware Relative Policy Optimization (SRPO)}
To address the above limitation, we propose Structure-aware Relative Policy Optimization (SRPO), which explicitly incorporates the structural discrepancy between sampled ranking permutations into policy optimization.
Instead of relying solely on scalar reward differences, SRPO jointly considers reward improvement and permutation-level ranking changes, enabling more structure-aware credit assignment for listwise ranking, as shown in Figure~\ref{fig:method}.

\subsubsection{Structure-normalized Preference} \label{subsubsec:pref} \hfill\newline
For any pair of sampled permutations $(L_i, L_j)$ in a group of ranking lists sampled from the current policy $\mathcal{G}=\{L_1,\ldots,L_G\}$, we first define their reward difference as $\Delta \mathcal{R}_{ij} = \mathcal{R}(L_i)-\mathcal{R}(L_j)$.
To quantify the structural discrepancy between two ranking lists, we introduce a top-weighted Kendall-tau distance:
\begin{equation} \begin{aligned} \label{eq:dis}
d(L_i,L_j)=& \sum_{1\leq a<b \leq k}  w_aw_b \cdot \\ 
& \mathbf{I}\big[
\left(\text{id}(L_i[a])>\text{id}(L_i[b])\right) \wedge
\left(\text{id}(L_j[a])<\text{id}(L_j[b])\right) \\  
& \vee
\left(\text{id}(L_i[a])<\text{id}(L_i[b])\right) \wedge
\left(\text{id}(L_j[a])>\text{id}(L_j[b])\right)
\big],
\end{aligned}\end{equation}
where $\text{id}(\cdot) \in \{1,...,k\}$ is the index value of elements in the original list after sorting, and the indicator function $\mathbf{I}(\cdot)$ denotes whether a pair of objects is ordered oppositely by $L_i$ and $L_j$~\cite{kendall1938new,deshwal2022bayesian,jiao2015kendall}. $w$ is the position weight, for example, $w_a=\frac{1}{[\text{log}_2(a+1)]^{\eta}}$ where $\eta \geq 0$ is a position-weighted scale hyperparameter.
Note that when $\eta=0$, it degenerates into the ordinary Kendall-tau distance.
It emphasizes disagreements occurring at higher-ranked positions, which are generally more important in ranking systems.
Based on the reward and structural distance, we define the structure-normalized preference as:
\begin{equation}
    S_{ij} = \frac{\Delta \mathcal{R}_{ij}}{d(L_i,L_j)+\epsilon},
\end{equation}
where $\epsilon$ is a small constant for numerical stability.
Intuitively, a large reward improvement achieved with only minor structural changes indicates a highly valuable local ranking refinement.
Conversely, reward gains obtained through drastic permutation changes are down-weighted.
Therefore, $S_{ij}$ captures the efficiency of reward improvement w.r.t. the structural modification in the ranking space.

\subsubsection{Contrastive Relative Advantage} \label{subsubsec:adv} \hfill\newline 
Based on the structure-normalized preference $S_{ij}$, we construct a contrastive advantage estimator by aggregating pairwise comparisons within the sampled group.
Specifically, the advantage of permutation $L_i$ is defined as
\begin{equation} \label{eq:adv}
    A_i^{RD}=\frac{1}{G-1}\sum_{j\neq i}\tanh\left(\frac{\alpha}{\text{std}\left(\{\mathcal{R}(L_i)\}_{i=1}^G\right)} S_{ij}\right),
\end{equation}
where $\alpha$ is a temperature parameter controlling the sharpness of pairwise preference comparisons; we set it to 1 in this paper.
Compared with the group-normalized advantage adopted in GRPO, it explicitly models the relative structural superiority of each sampled permutation.
When a ranking list consistently achieves better rewards than its neighboring permutations with only small structural perturbations, it receives a larger positive advantage.
In contrast, reward improvements that originate from excessive ranking perturbations are naturally suppressed through the distance normalization term.
The properties of this formula are discussed in detail in Appendix~\ref{app:formula}.

\subsubsection{Action-level Policy Optimization} \label{subsubsec:actionpolicy} \hfill\newline 
Unlike \cite{singh2019policy,tu2022reinforcement,gao2023policy} that optimize the complete ranking permutation as a single action, SRPO adopts an action-level optimization strategy.
It is analogous to token-level optimization in RL for LLMs~\cite{zheng2025group}, where each ranking decision contributes to the final policy update.
For a sampled ranking list $L_i$, let $L_{i,t}$ denote the candidate selected at ranking position $t$, and $L_{i,<t}$ denote the partial ranking prefix before position $t$.
The probability ratio is defined as $r_{\theta}(L_{i,t}\mid X, L_{i,<t})=\frac{\pi_\theta(L_{i,t}\mid X,L_{i,<t})}{\pi_{\theta_{\text{old}}}(L_{i,t}\mid X,L_{i,<t})}$, and the final SRPO objective is:
\begin{equation}\begin{aligned}
J_{\text{SRPO}}(\theta) = \mathbb{E}_{L_i \sim \pi_{\theta}} \frac{1}{G}\sum_{i=1}^G  \frac{1}{k} \sum_{t=1}^k 
\big[
r_{\theta}(L_{i,t}\mid & X,  L_{i,<t}) \cdot A_i^{RD} \\ - 
\beta & \cdot \mathbb{D}_{\mathrm{KL}}\left(\pi_\theta || \pi_{\mathrm{ref}}\right)
\big].
\end{aligned}\end{equation}

Compared with GRPO, SRPO introduces structure-aware pairwise preference modeling while performing policy optimization at the action level.
As a result, the learned policy is encouraged not only to maximize reward, but also to discover efficient local ranking improvements in the permutation space. 
Overall, SRPO provides a more structure-aware and stable credit assignment mechanism for listwise ranking. 
By measuring reward improvement per unit ranking discrepancy, it encourages efficient local refinements in the permutation space, while the pairwise aggregation and bounded nonlinear transformation reduce variance and suppress overly aggressive policy updates.

%% file: secs/4experi.tex
\input{tabs/ndcg_main}

\section{Experimental Setup}
To comprehensively evaluate the effectiveness of SRPO, we conduct experiments under two different ranking scenarios: learning to rank (LTR) and LLM-based text reranking.  
In particular, for LTR scenarios, we test both classical relevance optimization and more complicated fairness optimization. 
The relevance setting focuses on conventional ranking objectives such as NDCG, in which models are trained and evaluated using relevance-oriented rewards derived from ranking quality metrics. 
The fairness setting considers a more challenging ranking objective that aims to optimize exposure fairness across candidates while maintaining ranking utility.

\subsection{Datasets and Evaluation}
For LLM-based text rerankers, following previous works~\cite{liu2025e2rank,ke2026resrank}, we evaluate on both in-domain (TREC DL 19 and 20 datasets~\cite{craswell2025overview}) and out-of-domain (7 datasets of BEIR~\cite{thakur2021beir}) benchmarks, reporting the NDCG@10 metric.
We adopt the training data of E2Rank~\cite{liu2025e2rank}, which consists of 87k queries, each associated with 16 documents.

For LTR, we conduct experiments on three widely used LTR benchmarks: MSLR-WEB30K Fold 1 (MSLR), Istella-S LETOR (Istella), and Yahoo! LETOR Challenge set 1 (Yahoo)~\cite{qin2013introducing,chapelle2011yahoo,lucchese2016post}.
Each dataset consists of multiple query-specific candidate sets, in which each query-document pair is represented by a fixed-dimensional feature vector and an associated 5-level relevance (i.e., from 0 to 4; more detailed characteristics in Appendix~\ref{app:ltr_data}).
For the relevance experiments in LTR settings, we evaluate ranking quality on the test set using standard metrics, i.e., NDCG@\{3,10\} and ERR@\{3,10\}.
During evaluation, candidates are ranked deterministically based on the scores produced by the ranking model.
For the fairness experiments, we adopt the widely used individual exposure fairness metric, $\mathrm{fair}(\cdot)$.
Please see Appendix~\ref{app:ltr_data} for how it is calculated.

\input{tabs/rerank}

\subsection{Baselines}
We compare SRPO against representative supervised learning (SL) and reinforcement learning (RL) algorithms with both pointwise and listwise signals.
Specifically, these baselines are (see Appendix~\ref{app:baselines} for calculation details):
\begin{itemize}[leftmargin=*]
\item \textbf{CrossEntropy}~\cite{ai2021unbiased,burges2006learning}: An \textit{SL} ranking method that optimizes the cross-entropy loss between predicted ranking scores and normalized relevance labels.
\item \textbf{AttentionRank}~\cite{pang2020setrank,ai2018learning}: An \textit{SL} method that leverages attention-based supervision to emphasize highly relevant candidates.
\item \textbf{LambdaRank}~\cite{burges2010ranknet}: A \textit{SL} LTR algorithm that directly optimizes ranking metrics through lambda gradients derived from pairwise document comparisons.
\item \textbf{PLRank}~\cite{oosterhuis2021computationally,oosterhuis2022learning}: An \textit{RL} approach that utilizes decomposed \textit{item-level} rewards for policy optimization. We adopt an intuitive version, PLRank-0 (i.e., Eq (14) in~\cite{oosterhuis2021computationally}), as well as the most efficient vectorized version, PLRank-3~\cite{oosterhuis2022learning}.
\item \textbf{PGRank}~\cite{singh2019policy}: A policy-gradient-based \textit{RL} ranking method that directly optimizes listwise rewards by sampling ranking permutations from the Plackett-Luce distribution. Note that PGRank can be considered a \textit{listwise} version of PLRank.
\item \textbf{PPG}~\cite{xu2020reinforcement}: A pairwise policy-gradient \textit{RL} method that reduces gradient variance through intra-query comparisons between different samples. Following \citet{tu2022reinforcement}, we adopt a list-level reward formulation.
\item \textbf{GRPO}~\cite{shao2024deepseekmath,guo2025deepseek}: Group-relative policy optimization, a recent \textit{RL} algorithm that constructs advantages through reward normalization over multiple sampled rankings and has demonstrated strong performance in LLM post-training~\cite{guo2025deepseek} and LTR~\cite{tu2022reinforcement}.
\end{itemize} 

In LTR settings, following~\cite{tu2022reinforcement,tran2021ultra}, we adopt an MLP-based backbone model. 
For reranking, we select E2Rank~\cite{liu2025e2rank}, an efficient embedding-based ranking model based on Qwen3~\cite{yang2025qwen3}. 
The implementation details and training settings are detailed in Appendix~\ref{app:impl}.

%% file: tabs/ndcg_main.tex
\begin{table*}[htbp]
\caption{Test performance on Istella and Yahoo. The best results for different group size settings are bolded. "$^*$" indicates significantly better than the best supervised learning method (underlined) at the $p < 0.05$ level using the two-tailed pairwise t-test. "P" and "L" denote whether the algorithm uses pointwise labels or listwise rewards, respectively. Results on the MSLR dataset are reported in Appendix~\ref{app:exten_main_result} and Table~\ref{tab:ndcg_mslr}.}
\resizebox{0.99\textwidth}{!}{
\begin{tabular}{cccccccccc}
\toprule
\multicolumn{2}{c|}{Dataset} 
& \multicolumn{4}{c|}{Istella} 
& \multicolumn{4}{c}{Yahoo} \\
\midrule
\multicolumn{1}{c|}{Type} 
& \multicolumn{1}{c|}{Algorithm} 
& ERR@3 & ERR@10 & NDCG@3 & \multicolumn{1}{c|}{NDCG@10}
& ERR@3 & ERR@10 & NDCG@3 & NDCG@10 \\ 
\midrule

\multicolumn{10}{c}{Supervised Learning} \\ 
\midrule
\multicolumn{1}{c|}{\multirow{3}{*}{P}} 
& \multicolumn{1}{c|}{CrossEntropy} 
& 0.6928 & 0.7166 & 0.6224 & \multicolumn{1}{c|}{\underline{0.7088}}
& 0.4238 & 0.4610 & 0.6890 & 0.7580 \\
\multicolumn{1}{c|}{} 
& \multicolumn{1}{c|}{AttentionRank} 
& 0.6848 & 0.7093 & 0.6077 & \multicolumn{1}{c|}{0.6900}
& 0.4276 & 0.4646 & 0.6832 & 0.7515 \\
\multicolumn{1}{c|}{} 
& \multicolumn{1}{c|}{LambdaRank} 
& \underline{0.6991} & \underline{0.7224} & \underline{0.6276} & \multicolumn{1}{c|}{0.7074}
& \underline{0.4304} & \underline{0.4666} & \underline{0.6983} & \underline{0.7627} \\ 

\midrule
\multicolumn{10}{c}{Reinforcement Learning, Group Size $G=8$} \\ 
\midrule
\multicolumn{1}{c|}{\multirow{2}{*}{P}} 
& \multicolumn{1}{c|}{PLRank-0} 
& 0.6851 & 0.7093 & 0.6081 & \multicolumn{1}{c|}{0.6817}
& 0.4305 & 0.4669 & 0.6922 & 0.7589 \\ 
\multicolumn{1}{c|}{} 
& \multicolumn{1}{c|}{PLRank-3} 
& 0.6877 & 0.7122 & 0.6107 & \multicolumn{1}{c|}{0.6859}
& 0.4312 & 0.4675 & 0.6948 & 0.7599 \\ 

\midrule
\multicolumn{1}{c|}{\multirow{4}{*}{L}} 
& \multicolumn{1}{c|}{PGRank} 
& 0.6995 & 0.7225 & 0.6245 & \multicolumn{1}{c|}{0.7017}
& 0.4311 & 0.4675 & 0.6967 & 0.7620 \\ 
\multicolumn{1}{c|}{} 
& \multicolumn{1}{c|}{PPG} 
& 0.6978 & 0.7212 & 0.6244 & \multicolumn{1}{c|}{0.7024}
& 0.4308 & 0.4673 & 0.6976 & 0.7623 \\ 
\multicolumn{1}{c|}{} 
& \multicolumn{1}{c|}{GRPO} 
& 0.7035$^*$ & 0.7264$^*$ & 0.6334$^*$ & \multicolumn{1}{c|}{0.7150$^*$}
& 0.4328 & 0.4689 & 0.7036$^*$ & 0.7671$^*$ \\ 
\multicolumn{1}{c|}{} 
& \multicolumn{1}{c|}{SRPO (ours)} 
& \textbf{0.7041}$^*$ & \textbf{0.7269}$^*$ & \textbf{0.6348}$^*$ & \multicolumn{1}{c|}{\textbf{0.7174}$^*$}
& \textbf{0.4344}$^*$ & \textbf{0.4706}$^*$ & \textbf{0.7063}$^*$ & \textbf{0.7707}$^*$ \\  

\midrule
\multicolumn{10}{c}{Reinforcement Learning, Group Size $G=64$} \\ 
\midrule
\multicolumn{1}{c|}{\multirow{2}{*}{P}} 
& \multicolumn{1}{c|}{PLRank-0}
& 0.6939 & 0.7177 & 0.6200 & \multicolumn{1}{c|}{0.6967}
& 0.4336$^*$ & 0.4697$^*$ & 0.7009 & 0.7653 \\ 
\multicolumn{1}{c|}{} 
& \multicolumn{1}{c|}{PLRank-3} 
& 0.6957 & 0.7193 & 0.6208 & \multicolumn{1}{c|}{0.6990}
& 0.4332 & 0.4695 & 0.7000 & 0.7651 \\  

\midrule
\multicolumn{1}{c|}{\multirow{4}{*}{L}} 
& \multicolumn{1}{c|}{PGRank} 
& 0.6983 & 0.7214 & 0.6265 & \multicolumn{1}{c|}{0.7044}
& 0.4304 & 0.4668 & 0.6942 & 0.7602 \\ 
\multicolumn{1}{c|}{} 
& \multicolumn{1}{c|}{PPG} 
& 0.6998 & 0.7229 & 0.6279 & \multicolumn{1}{c|}{0.7052}
& 0.4317 & 0.4682 & 0.6976 & 0.7626 \\ 
\multicolumn{1}{c|}{} 
& \multicolumn{1}{c|}{GRPO} 
& 0.7025 & 0.7253 & 0.6311$^*$ & \multicolumn{1}{c|}{0.7088}
& 0.4334$^*$ & 0.4693 & 0.7040$^*$ & 0.7672$^*$ \\ 
\multicolumn{1}{c|}{} 
& \multicolumn{1}{c|}{SRPO (ours)} 
& \textbf{0.7037}$^*$ & \textbf{0.7262}$^*$ & \textbf{0.6349}$^*$ & \multicolumn{1}{c|}{\textbf{0.7163}$^*$}
& \textbf{0.4338}$^*$ & \textbf{0.4702}$^*$ & \textbf{0.7046}$^*$ & \textbf{0.7699}$^*$ \\ 

\bottomrule
\end{tabular}
}
\label{tab:ndcg_main}
\end{table*}

%% file: tabs/rerank.tex
\begin{table*}[htbp]
\renewcommand{\arraystretch}{1.0}
\caption{Test NDCG@10 performance of LLM-based text reranker (i.e., E2Rank). The best and second-best results are marked in bold and underlined, respectively. "Vanilla" refers to the model trained with the original approach in \cite{liu2025e2rank} without RL.}
\resizebox{0.99\textwidth}{!}{
\begin{tabular}{l|cc|cccccccc}
\toprule
 \multicolumn{1}{c|}{Benchmark} & \multicolumn{2}{c|}{in-domain} & \multicolumn{8}{c}{out-of-domain (BEIR)} \\ 
 \midrule
 \multicolumn{1}{c|}{Algorithm}  & DL19 & DL20 & Covid & NFCorpus & Touche & DBPedia & SciFact &  Signal & \multicolumn{1}{c|}{News} & Avg. of 7 \\ 
 \midrule
Vanilla & 0.7059 & 0.6751 & \underline{0.8260} & 0.3831 & \textbf{0.4334} & 0.4232 & 0.7341 & \underline{0.3590} & \multicolumn{1}{c|}{\underline{0.5195}} & \underline{0.5255} \\
\midrule
+PLRank-0 & \textbf{0.7148} & 0.6944 & 0.8121 & 0.3853 & 0.3990 & 0.4169 & 0.7418 & 0.3492 & \multicolumn{1}{c|}{0.5126} & 0.5167 \\
+PLRank-3 & 0.7131 & \textbf{0.6948} & 0.8128 & \textbf{0.3890} & 0.3973 & 0.4203 & 0.7467 & 0.3442 & \multicolumn{1}{c|}{0.5170} & 0.5182 \\
+PGRank & 0.7127 & 0.6911 & 0.8076 & 0.3867 & 0.3966 & 0.4248 & 0.7446 & 0.3482 & \multicolumn{1}{c|}{0.5155} & 0.5177 \\
+PPG & 0.7132 & 0.6928 & 0.7878 & 0.3860 & 0.3855 & 0.4213 & \underline{0.7481} & 0.3443 & \multicolumn{1}{c|}{0.5123} & 0.5122 \\
+GRPO & 0.7108 & \underline{0.6945} & 0.7837 & 0.3858 & 0.3903 & \underline{0.4247} & \textbf{0.7497} & 0.3390 & \multicolumn{1}{c|}{0.5134} & 0.5124 \\
+SRPO (ours) & \underline{0.7140} & 0.6923 & \textbf{0.8278} & \underline{0.3870} & \underline{0.4306} & \textbf{0.4265} & 0.7462 & \textbf{0.3593} & \multicolumn{1}{c|}{\textbf{0.5228}} & \textbf{0.5286} \\
\bottomrule
\end{tabular}
}
\label{tab:rerank}
\end{table*}

%% file: secs/5result.tex
\begin{figure*}[t]
    \centering    
    \includegraphics[width=0.95\textwidth]{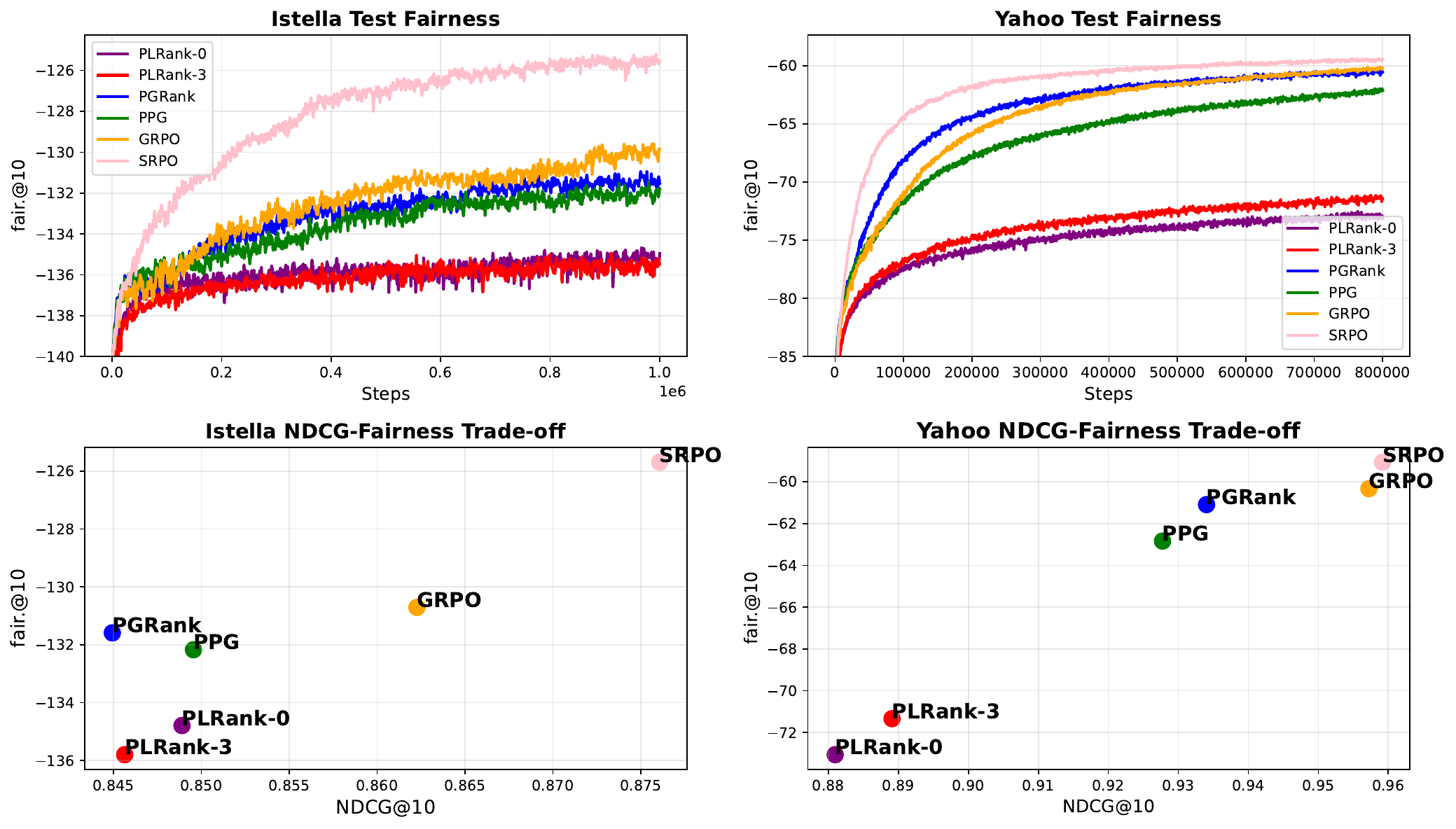}
    \caption{The fairness results and relevance-fairness trade-off of all RL algorithms.}
    \label{fig:fair}
\end{figure*}

\input{tabs/ablation}

\section{Results}
\subsection{Results of Ranking for Relevance}
Table~\ref{tab:ndcg_main} reports the relevance performance on Istella and Yahoo with NDCG@10 as the listwise reward and group size $G=8 / 64$, while results on MSLR and training curves are reported in Appendix~\ref{app:exten_main_result}. 
SRPO delivers the most consistent performance across datasets, metrics, and group sizes. 
Especially, on both Istella and Yahoo, it achieves the best results on all metrics and settings while matching or surpassing the strongest supervised baselines. 
This demonstrates that coarse-grained listwise rewards can provide effective supervision when paired with an appropriate policy optimization method.
Among RL methods, listwise approaches generally outperform PLRank, which relies on decomposed pointwise feedback, while GRPO improves over conventional policy-gradient methods through group-relative advantage estimation. 
SRPO further outperforms GRPO, showing that reward normalization alone cannot fully capture the differences among sampled permutations. 
Similar trends are observed when ERR@10 is used as the reward, as shown in Appendix~\ref{app:exten_main_result} and Table~\ref{tab:err}. 
Overall, by incorporating permutation-level discrepancy, SRPO emphasizes reward gains achieved through efficient structural changes. 
Moreover, $G=8$ generally matches or exceeds $G=64$ on Istella and Yahoo, indicating the strong performance of listwise RL methods under limited feedback (more analysis in \S\ref{subsubsec:groupsize}).

For reranking settings, Table~\ref{tab:rerank} reports the NDCG@10 performance of the E2Rank reranker with different training methods under in-domain and out-of-domain evaluation. 
Overall, RL-based training improves the in-domain results over the Vanilla model, while SRPO shows substantially better out-of-domain generalization. 
Although most RL baselines improve in-domain performance, their average BEIR scores decrease compared with the Vanilla model, except for SRPO. 
It suggests that directly optimizing scalar rewards may overfit domain-specific ranking patterns. 
By jointly considering reward improvements and permutation-level discrepancies, SRPO favors efficient local refinements and avoids overly aggressive reordering. 
Consequently, it improves in-domain effectiveness while better preserving the generalizable ranking ability.

\subsection{Results of Ranking for Fairness} 
Figure~\ref{fig:fair} reports the fairness performance and the relevance-fairness trade-off on Istella and Yahoo, where a larger $\mathrm{fair}@10$ value indicates a fairer allocation of exposure. 
As shown in the upper row, SRPO consistently achieves the highest fairness throughout the training stages and rapidly converges to the best final performance on both datasets. 
In the lower row, SRPO lies in the upper-right region on both datasets, achieving the best fairness while preserving the strongest NDCG, Pareto-dominating other methods. 
It improves both relevance and fairness over GRPO rather than trading one objective for the other, demonstrating a more favorable relevance-fairness balance.
Among other methods, PLRank-0 and PLRank-3 perform worse than the listwise methods, likely because fairness is a global property of the ranking distribution and is difficult to capture using decomposed candidate-level signals. 
On the other hand, PGRank, PPG, and GRPO improve through direct listwise optimization and relative reward estimation, but still compare rankings mainly through scalar rewards. 
Compared with them, SRPO additionally models permutation-level discrepancies, enabling it to favor efficient exposure adjustments over disruptive reorderings.

\subsection{Ablation Study}
To investigate the contribution of each component in SRPO, we conduct the following ablation variants:
\begin{itemize}[leftmargin=*]
\item w/o PW (in \S\ref{subsubsec:pref}): Remove the position weights in Eq.~\eqref{eq:dis} by setting $w_a=1$ for all positions. The resulting distance degenerates into the standard Kendall-tau distance and treats disagreements at different ranking positions equally.

\item w/o tanh (in \S\ref{subsubsec:adv}): Remove the $\tanh(\cdot)$ transformation from Eq.~\eqref{eq:adv}. This variant directly aggregates the unbounded structure-normalized pairwise preferences and is used to examine the effect of bounding and smoothing the relative advantages.
\item w/o std (in \S\ref{subsubsec:adv}): Remove the group-wise reward std term from Eq.~\eqref{eq:adv}. This variant tests whether dynamically adjusting the scale of pairwise comparisons according to the reward dispersion within each sampled group is beneficial.
\item out std (in \S\ref{subsubsec:adv}): Move the reward std term outside the $\tanh$ function in Eq.~\eqref{eq:adv}. It examines whether the reward-scale normalization should be applied before the nonlinear transformation or used only to rescale the resulting advantages.
\item w SL (in \S\ref{subsubsec:actionpolicy}): Replace the action-level policy optimization with sequence-level optimization~\cite{tu2022reinforcement,zheng2025group}. Specifically, the probability ratio and KL divergence are computed over the probability of the complete list, treating each sampled list as a single action.
\end{itemize}

Table~\ref{tab:abl} reports the ablation results on Istella and Yahoo. 
Overall, the complete SRPO achieves the best performance in seven out of the eight datasets, group-size, and metric combinations, and obtains the highest NDCG@10 in all settings. 
Removing the position weights consistently degrades performance across both datasets and group sizes, confirming that ranking disagreements should not be treated uniformly: structural changes near the top of a ranking list provide more informative optimization signals than equivalent changes at lower positions.
The variants related to the contrastive advantage estimator lead to the largest performance degradation. 
Removing the $\tanh$ function significantly reduces the performance across the four settings, showing that directly using unbounded pairwise preferences makes the optimization more sensitive to extreme reward differences or structurally anomalous samples, especially when there is insufficient feedback and the sample size is small. 
Removing the std term also causes consistent decreases in NDCG@10, demonstrating the benefit of adapting the comparison scale to the reward dispersion within each sampled group. 
Moreover, placing the std term outside $\tanh(\cdot)$ performs worse than the original formulation in all settings. 
This suggests that normalization should be applied before the bounded nonlinear transformation, so that the saturation level and discriminative strength of pairwise comparisons are jointly controlled by group-wise reward variation. 
Applying it only after $\tanh(\cdot)$ weakens this interaction and may no longer preserve a uniformly bounded advantage scale.
Finally, replacing action-level optimization with sequence-level optimization results in smaller but consistent performance reductions. 
This indicates that optimizing the probability ratios and KL regularization at individual ranking decisions provides finer-grained policy updates than treating the complete permutation as an atomic action. 
Nevertheless, the degradation is less pronounced than that caused by modifying the advantage estimator, suggesting that the primary gains of SRPO arise from its structure-aware credit assignment, while action-level optimization provides an additional and complementary improvement.

\begin{figure}[t]
    \centering
    \includegraphics[width=1.0\columnwidth]{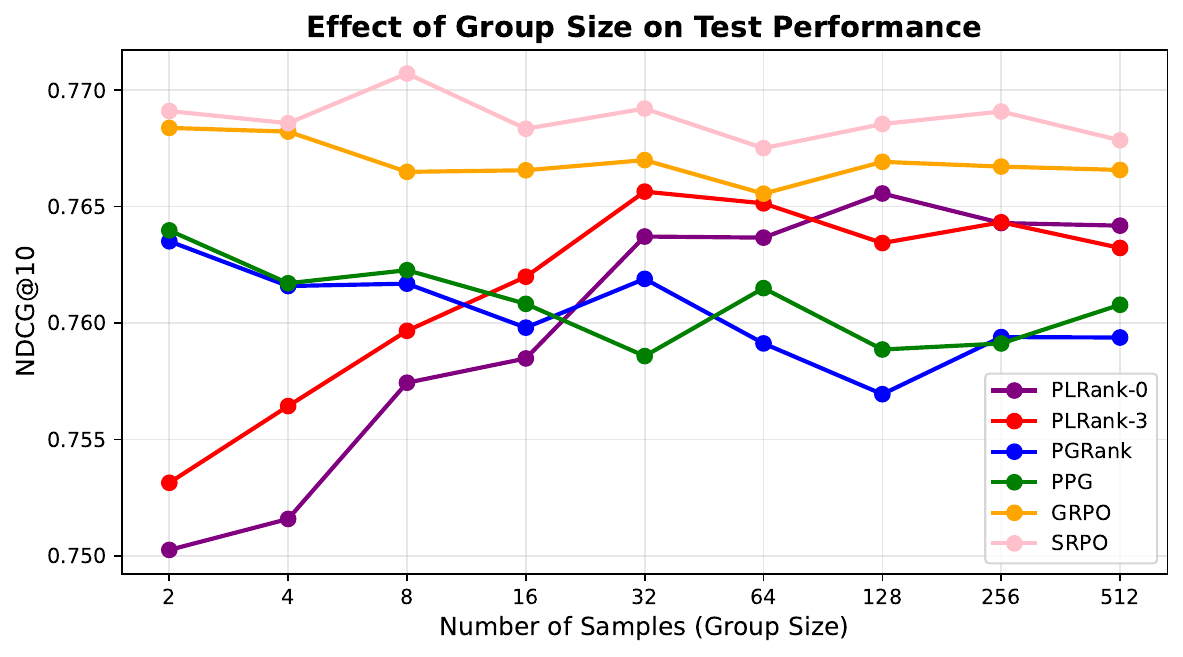}
    \caption{Effect of the group size $G$, i.e., the number of sampled lists for each query, on Yahoo test NDCG@10.}
    \label{fig:sample}
\end{figure}

\begin{figure}[t]
    \centering
    \includegraphics[width=1.0\columnwidth]{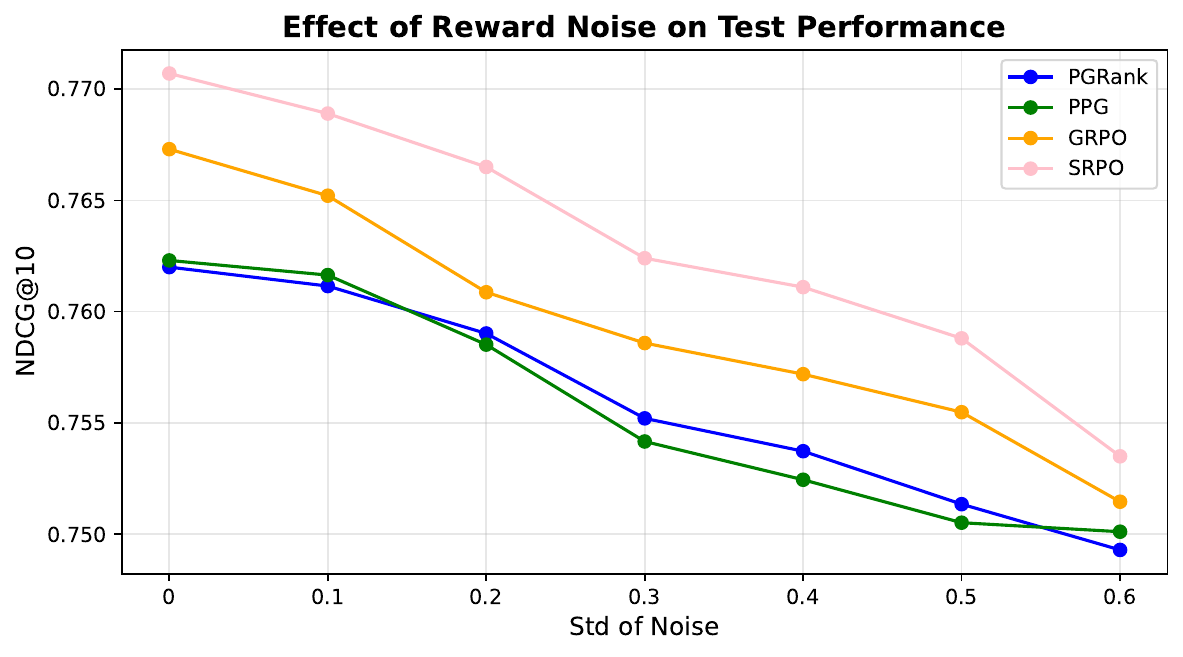}
    \caption{Effect of the reward noise on Yahoo test NDCG@10.}
    \label{fig:noise}
\end{figure}

\subsection{Analysis} 
\subsubsection{Group Size (Number of Samples)} \label{subsubsec:groupsize} \hfill \newline
The group size $G$ determines the number of ranking permutations sampled for each query during policy optimization. 
A larger group provides more comparisons for estimating relative advantages and may reduce the variance of policy updates, but it also incurs higher sampling and reward-evaluation costs. 
More importantly, real-world ranking systems often receive only limited repeated feedback for the same query. 
Therefore, an effective RL-based ranking method should not only benefit from richer sampling but also remain accurate and stable when only a small number of samples are available.
Figure~\ref{fig:sample} reports NDCG@10 on Yahoo as $G$ varies from 2 to 512. 
SRPO consistently achieves the best performance across all group sizes, demonstrating that its advantage is not restricted to a particular sampling configuration. 
In particular, SRPO already obtains strong performance with only two or four samples and reaches its highest result around $G=8$. 
This indicates that the structure-aware pairwise comparisons in SRPO can extract informative optimization signals from a small sampled group and are robust to changes in the amount of available feedback.
The remaining methods exhibit different sensitivities to $G$. 
PLRank-0 and PLRank-3 improve substantially as the group size increases from very small values, suggesting that their pointwise and candidate-level gradient estimates benefit from additional samples, but their performance eventually saturates. 
PGRank and PPG fluctuate more noticeably and do not show a monotonic improvement with larger groups, reflecting the instability of conventional policy-gradient estimation. 
GRPO is comparatively stable and consistently outperforms these baselines, confirming the benefit of group-relative reward normalization, though it remains weaker than SRPO.
These results suggest that incorporating permutation-level structure improves both sample efficiency and robustness: SRPO can construct reliable relative advantages under limited sampling, while additional samples are not required to maintain strong performance.

\subsubsection{Robustness to Noise} \hfill \newline
In practical ranking systems, coarse-grained feedback is typically derived from user interactions, such as clicks, dwell time, or session-level engagement. 
It may therefore be affected by behavioral stochasticity, observation noise, and other unobserved factors. 
It is thus important for an RL-based ranking method to remain effective when the observed listwise reward is noisy. 
To evaluate this property, we perturb the NDCG@10 reward with zero-mean Gaussian noise with std $\sigma$ ranging from 0 to 0.6 and report the test NDCG@10 on Yahoo. 
Since the noise is injected into the listwise reward, we focus on RL methods that directly optimize complete-ranking feedback.
As shown in Figure~\ref{fig:noise}, the performance of all methods gradually decreases as the noise level increases, confirming that inaccurate reward observations make listwise policy optimization more difficult. 
Nevertheless, SRPO consistently achieves the best test performance across the entire range of noise levels. 
Its advantage is especially clear under low and moderate noise, where it maintains a noticeable margin over other methods.
It indicates that structure-aware relative optimization is less susceptible to noisy list-level supervision.
These results support the applicability of SRPO to realistic ranking scenarios in which coarse-grained feedback may be imperfect.

%% file: tabs/ablation.tex
\begin{table*}[t]
\renewcommand{\arraystretch}{1.0}
\caption{Ablation experiment results. The best results are bolded. "$^\dagger$" indicates significantly worse than the original SRPO performance at the $p < 0.05$ level using the two-tailed pairwise t-test. "PW" and "SL" stand for position weight (in \S\ref{subsubsec:pref}) and sequential-level policy optimization (in \S\ref{subsubsec:actionpolicy}), respectively.}
\resizebox{0.99\textwidth}{!}{
\begin{tabular}{c|cccc|cccc}
\toprule
 Dataset & \multicolumn{4}{c|}{Istella} & \multicolumn{4}{c}{Yahoo} \\ 
 \midrule
 Group Size & \multicolumn{2}{c|}{$G$=8} & \multicolumn{2}{c|}{$G$=64} & \multicolumn{2}{c|}{$G$=8} & \multicolumn{2}{c}{$G$=64} \\
 \midrule
 Metrics& NDCG@3 & \multicolumn{1}{c|}{NDCG@10} & NDCG@3 & NDCG@10 & NDCG@3 & \multicolumn{1}{c|}{NDCG@10} & NDCG@3 & NDCG@10 \\
 \midrule
SRPO & 0.6348 & \multicolumn{1}{c|}{\textbf{0.7174}} & \textbf{0.6349} & \textbf{0.7163} & \textbf{0.7063} & \multicolumn{1}{c|}{\textbf{0.7707}} & \textbf{0.7046} & \textbf{0.7699} \\ 
\midrule
w/o PW (\S\ref{subsubsec:pref})& 0.6330 & \multicolumn{1}{c|}{0.7135} & 0.6310 & 0.7118$^\dagger$ & 0.7023 & \multicolumn{1}{c|}{0.7682} & 0.7005$^\dagger$ & 0.7662 \\
w/o tanh (\S\ref{subsubsec:adv}) & 0.6281$^\dagger$ & \multicolumn{1}{c|}{0.7111$^\dagger$} & 0.6346 & 0.7116$^\dagger$ & 0.6987$^\dagger$ & \multicolumn{1}{c|}{0.7642$^\dagger$} & 0.6993$^\dagger$ & 0.7649$^\dagger$ \\
w/o std (\S\ref{subsubsec:adv}) & \textbf{0.6350} & \multicolumn{1}{c|}{0.7124$^\dagger$} & 0.6346 & 0.7118$^\dagger$ & 0.7002$^\dagger$ & \multicolumn{1}{c|}{0.7657$^\dagger$} & 0.6997$^\dagger$ & 0.7646$^\dagger$ \\
out std (\S\ref{subsubsec:adv}) & 0.6283$^\dagger$ & \multicolumn{1}{c|}{0.7117$^\dagger$} & 0.6272$^\dagger$ & 0.7110$^\dagger$ & 0.6994$^\dagger$ & \multicolumn{1}{c|}{0.7644$^\dagger$} & 0.6990 & 0.7659 \\
w SL (\S\ref{subsubsec:actionpolicy}) & 0.6331 & \multicolumn{1}{c|}{0.7156} & 0.6313 & 0.7154 & 0.7051 & \multicolumn{1}{c|}{0.7692} & 0.7039 & 0.7682 \\
\bottomrule
\end{tabular}
}
\label{tab:abl}
\end{table*}

%% file: secs/6conclu.tex
\section{Conclusions}
In this paper, we identify a fundamental limitation of existing RL methods for ranking: they compare sampled ranking lists primarily through scalar rewards while overlooking the structural relationships among permutations. 
To address this issue, we propose SRPO, a structure-aware relative policy optimization framework for listwise ranking. 
It measures permutation discrepancy using a top-weighted Kendall-tau distance and normalizes pairwise reward differences by the corresponding structural changes, thereby capturing the reward improvement achieved per unit of ranking modification. 
Extensive experiments on LTR and LLM-based text reranking tasks demonstrate the effectiveness and generality of SRPO, showing that explicitly modeling structural differences among ranking lists enables more accurate and robust credit assignment for listwise policy optimization. 
Future work may extend SRPO to online interaction feedback, partial rankings, larger candidate spaces, and more complex multi-objective ranking scenarios.

%% file: secs/7appendix.tex
\appendix

\section{Efficent Plackett-Luce Sampling} \label{app:pl}
Direct sampling from the ranking score distribution can be computationally expensive.
Following prior works~\cite{oosterhuis2021computationally,oosterhuis2022learning,tu2022reinforcement}, we employ the Gumbel Softmax trick~\cite{gumbel1954statistical} for efficient sampling.
Specifically, independent Gumbel noise is added to each ranking score:
\begin{equation}
    \hat h_\theta(x_i)=h_\theta(x_i)+\gamma_i,
\end{equation}
where $\gamma_i \sim \mathrm{Gumbel}(0,0)$.
A sampled list is then obtained by sorting candidates according to the perturbed scores $\hat h_\theta(x_i)$ in descending order.
This procedure generates exact samples from the score distribution while reducing the sampling complexity to
$\mathcal{O}(k\log k)$.

\section{Discussion: Properties of the Contrastive Relative Advantage} \label{app:formula}
Our contrastive relative advantage (Eq.~\eqref{eq:adv}) provides the following desirable properties:
\begin{itemize}[leftmargin=*]
\item Improved Local Credit Assignment: Ranking improvements are attributed according to local structural modifications rather than global reward magnitudes. When a ranking list can earn a higher reward through minor adjustments to its sort order, especially in top slots, it is assigned a higher preference value. This property is desirable in ranking tasks, where local ordering changes, especially at top positions, often carry more meaningful optimization signals than arbitrary global permutation changes.
\item Variance Reduction: Pairwise aggregation reduces the influence of extreme rewards within a group. Since $S_{ij}=-S_{ji}$ and $\tanh(\cdot)$ is an odd function, the pairwise comparison terms cancel across the group, implying that $\sum_i A_i^{RD}=0$. This naturally centers the advantage estimator and reduces systematic update bias.  
\item Smoother Optimization Landscape: The advantage is bounded because $\tanh(\cdot)\in[-1,1]$. Hence, extreme reward differences or noisy samples cannot produce unbounded policy gradients, thus preventing overly aggressive updates caused by outlier reward differences. Moreover, since $\tanh(\cdot)$ is Lipschitz-continuous, small perturbations in rewards only lead to bounded changes in the advantage estimate. Therefore, it produces smoother advantage estimates under noisy reward signals.
\item Dynamic Controllability: The standard deviation term $\text{std}(\cdot)$ can maintain the stability of training. When the variance of rewards within a group increases, the group's reward decreases, thereby reducing instability in estimating advantages. Conversely, when rewards within a group are concentrated, the discriminatory power of pairwise comparisons is enhanced.
\end{itemize}

\input{tabs/dataset}

\section{LTR Datasets and Fairness Evaluation} \label{app:ltr_data}
The detailed properties of the three LTR datasets are outlined in Table~\ref{tab:data}. 
We follow the data preprocessing pipeline in~\cite{tran2021ultra}.

The fairness metric, $\mathrm{fair}(\cdot)$, measures the consistency between document relevance $R$ and allocated exposure $E$~\cite{oosterhuis2021computationally,oosterhuis2022learning,tu2022reinforcement,yang2021maximizing}.
It is calculated as:
\begin{equation}\label{eq:fair}
\mathrm{fair}(q)=-\mathrm{unfair}(q),
\end{equation}
\begin{equation}
\mathrm{unfair}(q) = \frac{1}{k(k - 1)} \sum_{d_x, d_y \in L^q} \left(E(d_x)R(d_y) - E(d_y)R(d_x)\right)^2,
\end{equation}
\begin{equation}
E(d) = \sum_{\tau = 1}^T\sum_{i=1}^k\frac{1}{\log_2(i + 1)}\mathbf{I}\big(L_{\tau}^q[i] = d\big),  
\end{equation} 
\begin{equation}
    R(d) = \frac{2^{y_d} - 1}{2^{y_{\rm{max}}} - 1},
\end{equation}
where $L_{\tau}^q$ refers to the rank list at the $\tau$-th sampling w.r.t. query $q$, $y_d$ represents the relevance judgment of $d$, and $y_{\rm{max}} = 4$ is the maximum relevance judgment of the dataset.
Following previous studies~\cite{tu2022reinforcement,oosterhuis2021computationally,oosterhuis2022learning}, exposure $E(\cdot)$ is estimated from repeatedly sampled ranking lists for $T=100$ times.

\section{Formulas for Baselines} \label{app:baselines}
Here, we provide a detailed explanation of the calculation process for each baseline:
\begin{itemize}[leftmargin=*]
    \item \textbf{CrossEntropy~\cite{ai2021unbiased,burges2006learning}} converts the model scores into a listwise probability distribution and minimizes the cross-entropy against normalized relevance labels:
    \begin{equation}
        a_i^h =
        \frac{\exp(h_\theta(x_i))}
        {\sum_{j=1}^{k}\exp(h_\theta(x_j))},
    \end{equation}
    \begin{equation}
        \mathcal{J}_{\mathrm{CrossEntropy}}(\theta)
        =
        -\sum_{i=1}^{k}
        \frac{y_i}{\sum_{j=1}^{k}y_j}
        \log a_i^h.
    \end{equation}
    This objective encourages the predicted score distribution to match the relative relevance distribution within each query.
    \item \textbf{AttentionRank~\cite{pang2020setrank,ai2018learning}} assigns greater supervision weight to highly relevant candidates by exponentially transforming positive relevance labels:
    \begin{equation}
        a_i^y =
        \begin{cases}
            \displaystyle
            \frac{\exp(y_i)}
            {\sum_{j:y_j>0}\exp(y_j)},
            & y_i>0, \\
            0, & \text{otherwise}.
        \end{cases}
    \end{equation}
    It then applies a candidate-wise binary cross-entropy objective:
    \begin{equation}
        \mathcal{J}_{\mathrm{AttentionRank}}(\theta)
        =
        -\sum_{i=1}^{k}
        \left[
        a_i^y\log a_i^h
        +(1-a_i^y)\log(1-a_i^h)
        \right].
    \end{equation}

    \item \textbf{LambdaRank~\cite{burges2010ranknet}} optimizes pairwise score differences while weighting each pair by its potential impact on the target ranking metric:
    \begin{equation}
    \begin{aligned}
        \mathcal{J}_{\mathrm{LambdaRank}}(\theta)
        = &
        \sum_{i=1}^{k}
        \sum_{j:y_j<y_i}
        \Delta\mathrm{Rel}(i,j) \ \cdot \\ 
        & \log_2
        \left(
        1+
        \exp\left[
        -\sigma
        \big(
        h_\theta(x_i)-h_\theta(x_j)
        \big)
        \right]
        \right),
    \end{aligned}
    \end{equation}
    where $\Delta\mathrm{Rel}(i,j)$ is the absolute change in the target metric, such as NDCG, caused by exchanging candidates $i$ and $j$.

    \item \textbf{PLRank~\cite{oosterhuis2021computationally,
    oosterhuis2022learning}}. Unlike methods trained only with a scalar listwise reward, PLRank assumes that the ranking utility can be decomposed into candidate- and position-level contributions. It assigns each sequential PL decision the accumulated reward of the remaining positions:
    \begin{equation}
    \begin{aligned}
        \nabla_\theta J_{\mathrm{PLRank}}(\theta)
        =
        \mathbb{E}_{L\sim\pi_\theta} & 
        \bigg[
        \sum_{i=1}^{k}
        \sum_{j=i}^{k}
        \mathcal{R}(L[j]) \ \cdot \\ 
        & \nabla_\theta
        \log\pi_\theta
        \big(
        L[i]\mid X,L[1:i-1]
        \big) 
        \bigg],
    \end{aligned}
    \end{equation}
    where $\mathcal{R}(L[j])$ denotes the decomposed reward contribution of the candidate placed at position $j$. We denote this direct estimator as \textbf{PLRank-0}. \textbf{PLRank-3} is its computationally optimized and vectorized implementation proposed by \citet{oosterhuis2022learning}.

    \item \textbf{PGRank~\cite{singh2019policy}} treats a complete ranking permutation as one stochastic action and directly maximizes its expected listwise reward using REINFORCE:
    \begin{equation}
    \begin{aligned}
        \nabla_\theta J_{\mathrm{PGRank}}(\theta)
        =
        \mathbb{E}_{L\sim\pi_\theta}
        \left[
        \big(
        \mathcal{R}(L)-b(q)
        \big)
        \nabla_\theta
        \log\pi_\theta(L\mid X)
        \right],
    \end{aligned}
    \label{eq:pgrank}
    \end{equation}
    where $b(q)$ is the average reward of multiple rankings sampled for query $q$. The baseline reduces gradient variance without changing the expected policy gradient.

    \item \textbf{PPG~\cite{xu2020reinforcement}} constructs a relative training signal from two permutations sampled for the same query. In our listwise adaptation, the gradient is determined by both their reward difference and their log-probability difference:
    \begin{equation}
    \begin{aligned}
        \nabla_\theta J_{\mathrm{PPG}}(\theta)
        =
        & \mathbb{E}_{L_1,L_2\sim\pi_\theta}
        \Big[
        \big(
        \mathcal{R}(L_1)-\mathcal{R}(L_2)
        \big) \cdot
        \\
        &
        \big(
        \nabla_\theta\log\pi_\theta(L_1\mid X)
        -
        \nabla_\theta\log\pi_\theta(L_2\mid X)
        \big)
        \Big].
    \end{aligned}
    \end{equation}
    Thus, each sampled ranking implicitly serves as a baseline for the other, reducing dependence on absolute reward magnitudes.
    
    \item  \textbf{GRPO~\cite{guo2025deepseek,shao2024deepseekmath}}. See \S\ref{subsubsec:grpo}.
\end{itemize}

\input{tabs/ndcg_mslr}

\begin{figure*}[t]
    \centering    
    \includegraphics[width=0.9\textwidth]{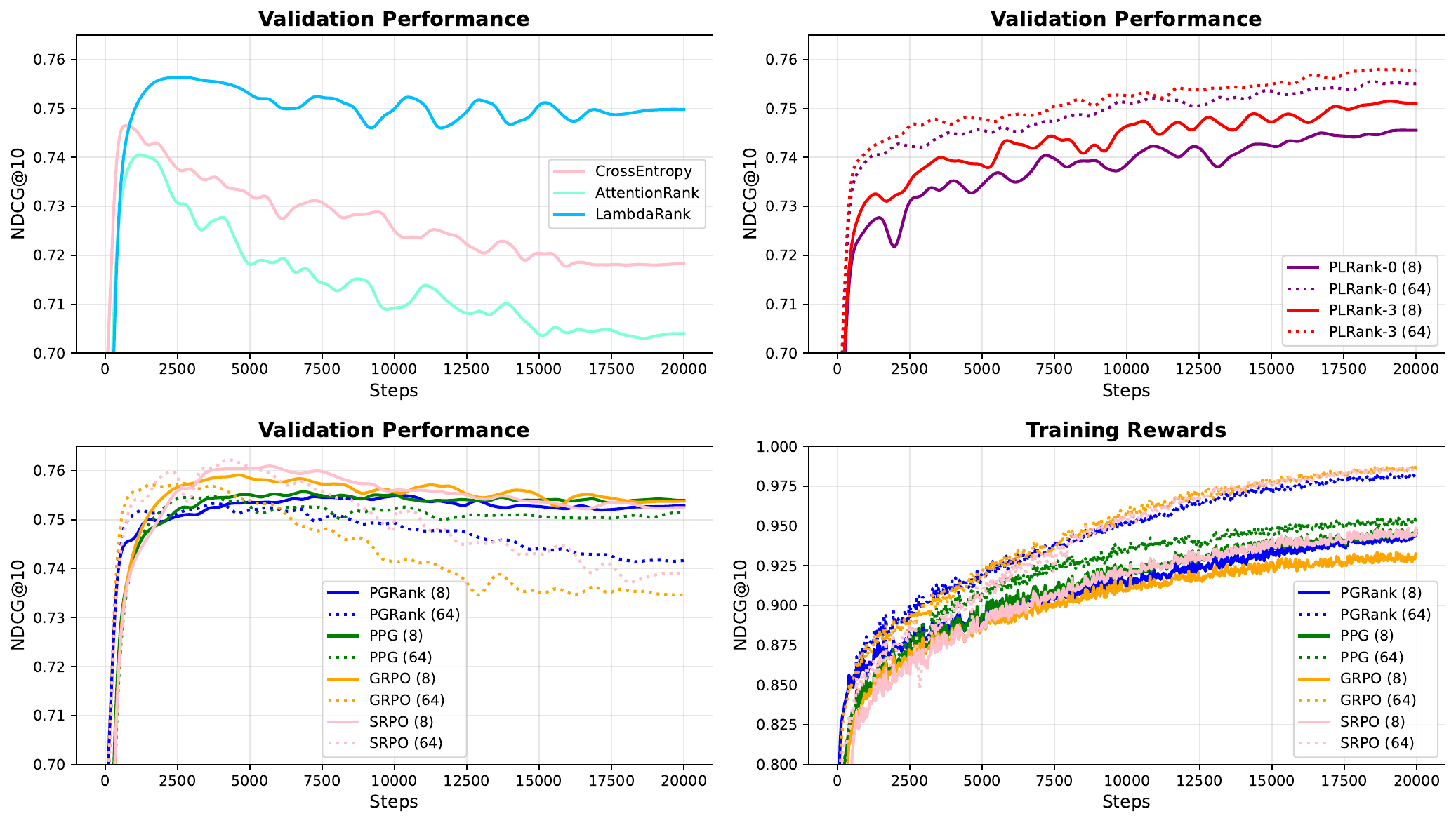}
    \caption{Validation curves of all algorithms and reward trajectories of listwise RL methods on Yahoo.}
    \label{fig:curve}
\end{figure*}

\input{tabs/err}

\section{Implementation Details} \label{app:impl}
\subsection{Settings of LTR}
\subsubsection{Reward Simulation} \hfill \newline
Since real-world search engine result page (SERP) feedback is unavailable in standard LTR benchmarks, we follow prior works~\cite{oosterhuis2018ranking,tu2022reinforcement} to simulate listwise rewards from candidate-level relevance annotations. 
In the relevance experiments, the reward of a sampled ranking list is computed by applying a target relevance metric, such as NDCG@10 or ERR@10, to the top-ranked results. 
This simulates session-level feedback that reflects the overall quality of the displayed ranking list. 
In the fairness experiments, we adopt the exposure-gradient-based reward $\mathcal{R}^{\mathrm{fair}}$ used in previous fairness-aware ranking studies~\cite{tu2022reinforcement,oosterhuis2021computationally}, which encourages the model to allocate exposure more consistently with candidate relevance:
\begin{equation}
\mathcal{R}^{\mathrm{fair}}(L) = \sum_{i=1}^k \frac{\rho^{\mathrm{fair}}_{L[i]}}{\log_2(i + 1)},
\end{equation}
\begin{equation}\begin{aligned}
\rho^{\mathrm{fair}}_d & = \frac{\partial \ \mathrm{fair}(q)}{\partial \ E(d)}  \\ & =  \frac{4}{k(k-1)}  \sum_{d' \in L^q} \left(E(d')R(d) - E(d)R(d')\right)R(d').
\end{aligned}\end{equation}

\subsubsection{Training Settings} \hfill \newline
We simulate an online ranking environment in which a user only observes the top ($k=10$) candidates returned by the ranking model and provides a list-level feedback signal on these observed candidates.
To ensure fair comparison, for supervised learning baselines, models are also optimized on these top candidates based on ground-truth labels.
For RL-based methods, multiple ranking lists are sampled for each query to estimate policy gradients or relative advantages.
We evaluate two group-size settings, $G=8$ and $G=64$, corresponding to low-feedback and high-feedback regimes, respectively. 
The former better reflects realistic ranking scenarios where repeated feedback for the same query is limited, while the latter approximates a richer-feedback setting.
For relevance optimization, all models are trained with a batch size of 256 using AdamW for 20,000 steps.
The learning rate of all methods is selected from the range $[10^{-5}, 10^{-2}]$ via grid search based on the best validation performance.
The checkpoint with the best validation performance is selected for test evaluation.
For SRPO, the range of values for each hyperparameter is: KL penalty weight $\beta \in [0, 0.1]$, position scale $\eta \in \{1, 2, 3\}$.
The reference policy is periodically updated every 500 steps during training, following \cite{tu2022reinforcement}. 

On the other hand, in fairness settings, due to the higher variance and complexity of the fairness reward, following~\cite{oosterhuis2021computationally,tu2022reinforcement}, we adopt a larger sampling size $G=100$ and a relatively high learning rate of 1e-3.
Additionally, we increase the number of training steps until all algorithms converge.
For each query, we only keep the top 10 candidates provided by SVM-rank to ensure a consistent dimension $|L^q|=k=10$ for calculating the fairness in Eq.~\eqref{eq:fair}, following~\cite{tran2021ultra,tu2022reinforcement}. 
As a result, since the candidate set is pre-truncated, the value of relevance metrics in fairness experiments is not directly comparable to that in relevance-focused experiments on the same dataset.
All LTR experiments are conducted on a single \textit{NVIDIA GeForce RTX 3090} GPU.

\subsection{Settings of LLM Reranker (E2Rank)}
We adopt E2Rank-0.6B~\cite{liu2025e2rank} as the backbone for LLM-based text reranking and closely follow its Stage II continued training pipeline\footnote{\url{https://github.com/Alibaba-NLP/E2Rank}}. 
The only change is to the training objective, where we add an additional RL loss term, including PLRank, PGRank, PPG, GRPO, or our SRPO, with group size $G=64$.
We also adopt NDCG@10 as the reward function.
The parameter settings remain consistent with the E2Rank original paper, and the model trained using the original approach is referred to as \textit{Vanilla} in Table~\ref{tab:rerank}.
Starting from the embedding checkpoint\footnote{\url{https://huggingface.co/Alibaba-NLP/E2Rank-0.6B-Embedding-Only}} obtained after Stage I, each training instance contains one query, one positive document, and 15 hard negatives. 
The query and all 16 candidate documents are concatenated into a listwise prompt, which is encoded as a pseudo-relevance-feedback-enhanced query.
The reranking score of candidate $d_i$ is computed as the cosine similarity between its document embedding and the embedding of the listwise prompt.
This formulation allows the model to capture query-document and document-document interactions through the listwise context while retaining the efficient similarity-based scoring mechanism of an embedding model.

\section{Extensive LTR Relevance Results} \label{app:exten_main_result}
\subsection{Results on MSLR}
Table~\ref{tab:ndcg_mslr} demonstrates the LTR performance of all algorithms on the MSLR dataset, serving as a supplement to Table~\ref{tab:ndcg_main}.
It shows a similar pattern to Table~\ref{tab:ndcg_main}.
With $G=8$, SRPO achieves the best NDCG@3 and NDCG@10, although GRPO is marginally better on the ERR metrics. 
When the group size increases to 64, SRPO obtains the best result on all four evaluation metrics.

\subsection{Reward \& Validation Curves}
On the other hand, Figure~\ref{fig:curve} shows the training dynamics on Yahoo. 
The supervised methods converge quickly, but CrossEntropy and AttentionRank gradually overfit, while LambdaRank remains relatively stable. 
PLRank-0 and PLRank-3 converge more slowly and benefit clearly from increasing the group size from $G=8$ to $G=64$.
The listwise RL methods, especially GRPO and SRPO, reach stronger validation performance, with SRPO achieving the highest peak NDCG@10 and GRPO following closely. 
Notably, larger group sizes do not consistently improve validation performance and may lead to stronger late-stage degradation. 
Although the training rewards of all listwise methods increase steadily, these gains do not always translate into better validation results, indicating possible over-optimization of the sampled reward.
Overall, SRPO provides the most effective validation improvement, while the results also highlight the importance of validation-based checkpoint selection.

\subsection{ERR as Reward}
We further examine robustness to the choice of reward function by replacing NDCG@10 with ERR@10, as reported in Table~\ref{tab:err}. 
SRPO still achieves the best performance across all metrics and group sizes on Istella and Yahoo. 
On MSLR, it remains competitive on the directly optimized ERR metrics while consistently obtaining the best NDCG performance. 
It suggests that structure-aware optimization does not merely overfit the scalar training reward; instead, it encourages ranking modifications that transfer well across different relevance metrics.

%% file: tabs/dataset.tex
\begin{table}[t]
    \centering
    \caption{A summary of the dataset statistics.}
    \renewcommand{\arraystretch}{1.1}
    \begin{tabular}{c|ccc}
    \toprule
    Property/Dataset & MSLR & Istella & Yahoo  \\
    \midrule
    \#Queries & 31,151 & 33,018 & 29,921   \\
    \#Candidates & 376k & 340k & 71k  \\
    \#Avg. candidates per query & 121 & 103 & 24  \\
    \#Feature dimension & 136 & 220 & 700  \\
    \#Label level & 5 & 5 & 5 \\  
    \bottomrule
    \end{tabular}
    \label{tab:data}
\end{table}

%% file: tabs/ndcg_mslr.tex
\begin{table}[t]
\renewcommand{\arraystretch}{1.2}
\caption{Test performance on MSLR. The best results for different group size settings are bolded. "$^*$" indicates significantly better than the best supervised learning method (underlined) at the $p < 0.05$ level using the two-tailed pairwise t-test. "P" and "L" denote whether the algorithm uses pointwise labels or listwise rewards, respectively.}
\label{tab:ndcg_mslr}
\resizebox{1.0\columnwidth}{!}{
\begin{tabular}{cccccc}
\toprule
\multicolumn{1}{c|}{Type} & \multicolumn{1}{c|}{Algorithm} & ERR@3 & ERR@10 & NDCG@3 & NDCG@10 \\ 
\midrule
\multicolumn{6}{c}{Supervised Learning} \\ 
\midrule
\multicolumn{1}{c|}{\multirow{3}{*}{P}} & \multicolumn{1}{c|}{CrossEntropy} 
& 0.3031 & 0.3457 & 0.4242 & 0.4466 \\
\multicolumn{1}{c|}{} & \multicolumn{1}{c|}{AttentionRank} 
& 0.3054 & 0.3474 & 0.4157 & 0.4369 \\
\multicolumn{1}{c|}{} & \multicolumn{1}{c|}{LambdaRank} 
& \underline{0.3124} & \underline{0.3540} & \underline{0.4313} & \underline{0.4521} \\
\midrule
\multicolumn{6}{c}{Reinforcement Learning, Group Size $G=8$} \\ 
\midrule
\multicolumn{1}{c|}{\multirow{2}{*}{P}} & \multicolumn{1}{c|}{PLRank-0} 
& 0.2962 & 0.3379 & 0.3981 & 0.4225 \\
\multicolumn{1}{c|}{} & \multicolumn{1}{c|}{PLRank-3} 
& 0.2965 & 0.3383 & 0.4016 & 0.4252 \\
\midrule
\multicolumn{1}{c|}{\multirow{4}{*}{L}} & \multicolumn{1}{c|}{PGRank} 
& 0.3107 & 0.3526 & 0.4270 & 0.4466 \\
\multicolumn{1}{c|}{} & \multicolumn{1}{c|}{PPG} 
& 0.3088 & 0.3510 & 0.4233 & 0.4451 \\
\multicolumn{1}{c|}{} & \multicolumn{1}{c|}{GRPO} 
& \textbf{0.3110} & \textbf{0.3543} & 0.4296 & 0.4520 \\
\multicolumn{1}{c|}{} & \multicolumn{1}{c|}{SRPO(ours)} 
& 0.3102 & 0.3535 & \textbf{0.4304} & \textbf{0.4530} \\  
\midrule
\multicolumn{6}{c}{Reinforcement Learning, Group Size $G=64$} \\ 
\midrule
\multicolumn{1}{c|}{\multirow{2}{*}{P}} & \multicolumn{1}{c|}{PLRank-0}
& 0.3072 & 0.3486 & 0.4124 & 0.4338 \\
\multicolumn{1}{c|}{} & \multicolumn{1}{c|}{PLRank-3} 
& 0.3081 & 0.3497 & 0.4154 & 0.4373 \\   
\midrule
\multicolumn{1}{c|}{\multirow{4}{*}{L}} & \multicolumn{1}{c|}{PGRank} 
& 0.3137 & 0.3556 & 0.4292 & 0.4492 \\
\multicolumn{1}{c|}{} & \multicolumn{1}{c|}{PPG} 
& 0.3130 & 0.3549 & 0.4290 & 0.4492 \\
\multicolumn{1}{c|}{} & \multicolumn{1}{c|}{GRPO} 
& 0.3142 & 0.3565$^*$ & 0.4332 & 0.4543 \\
\multicolumn{1}{c|}{} & \multicolumn{1}{c|}{SRPO (ours)} 
& \textbf{0.3151}$^*$ & \textbf{0.3571}$^*$ & \textbf{0.4366}$^*$ & \textbf{0.4564}$^*$\\ 
\bottomrule
\end{tabular}
}
\end{table}

%% file: tabs/err.tex
\begin{table}[t]
\centering
\caption{Test performance when using ERR@10 as the reward function. The best results for different group size settings are bolded. "$^\dagger$" indicates significantly worse than the best method with the same group size at the $p < 0.05$ level using the two-tailed pairwise t-test.}
\renewcommand{\arraystretch}{1.2}
\label{tab:err}
\resizebox{1.0\columnwidth}{!}{
\begin{tabular}{c|c|c||cccc}
\toprule
Dataset & $G$ & Algorithm & ERR@3 & ERR@10 & NDCG@3 & NDCG@10 \\ 
\midrule
\multirow{8}{*}{MSLR} & \multirow{4}{*}{8} 
    & PGRank & 0.3112 & 0.3522 & 0.4182$^\dagger$ & 0.4387$^\dagger$ \\
 &  & PPG & 0.3089 & 0.3502 & 0.4147$^\dagger$ & 0.4354$^\dagger$ \\
 &  & GRPO & \textbf{0.3117} & \textbf{0.3526} & 0.4278 & 0.4496 \\
 &  & SRPO & 0.3101 & 0.3509 & \textbf{0.4285} & \textbf{0.4511} \\ 
 \cline{2-7} 
 & \multirow{4}{*}{64} 
    & PGRank & 0.3132 & 0.3527 & 0.4200$^\dagger$ & 0.4420$^\dagger$ \\
 &  & PPG & \textbf{0.3141} & \textbf{0.3552} & 0.4218$^\dagger$ & 0.4417$^\dagger$ \\
 &  & GRPO & 0.3129 & 0.3547 & 0.4323 & 0.4534 \\
 &  & SRPO & 0.3132 & 0.3542 & \textbf{0.4324} & \textbf{0.4539} \\ 
 \midrule
\multirow{8}{*}{Istella} & \multirow{4}{*}{8} 
    & PGRank & 0.6937$^\dagger$ & 0.7173$^\dagger$ & 0.6158$^\dagger$ & 0.6896$^\dagger$ \\
 &  & PPG & 0.6960$^\dagger$ & 0.7193$^\dagger$ & 0.6162$^\dagger$ & 0.6889$^\dagger$ \\
 &  & GRPO & 0.6999 & 0.7226 & 0.6304 & 0.7108 \\
 &  & SRPO & \textbf{0.7027} & \textbf{0.7250} & \textbf{0.6334} & \textbf{0.7123} \\ 
 \cline{2-7} 
 & \multirow{4}{*}{64} 
    & PGRank & 0.6921$^\dagger$ & 0.7159$^\dagger$ & 0.6134$^\dagger$ & 0.6881$^\dagger$ \\
 &  & PPG & 0.6921$^\dagger$ & 0.7156$^\dagger$ & 0.6151$^\dagger$ & 0.6890$^\dagger$ \\
 &  & GRPO & 0.6987 & 0.7224 & 0.6282$^\dagger$ & 0.7090$^\dagger$ \\
 &  & SRPO & \textbf{0.7018} & \textbf{0.7247} & \textbf{0.6322} & \textbf{0.7141} \\ 
 \midrule
\multirow{8}{*}{Yahoo} & \multirow{4}{*}{8} 
    & PGRank & 0.4320 & 0.4685 & 0.6927$^\dagger$ & 0.7580 $^\dagger$\\
 &  & PPG & 0.4314 & 0.4681 & 0.6937$^\dagger$ & 0.7597$^\dagger$ \\
 &  & GRPO & 0.4323 & 0.4686 & 0.7012 & 0.7655 \\
 &  & SRPO & \textbf{0.4334} & \textbf{0.4695} & \textbf{0.7040} & \textbf{0.7681} \\ 
 \cline{2-7} 
 & \multirow{4}{*}{64} 
    & PGRank & 0.4312 & 0.4677 & 0.6911$^\dagger$ & 0.7568$^\dagger$ \\
 &  & PPG & 0.4312 & 0.4676 & 0.6919$^\dagger$ & 0.7575$^\dagger$ \\
 &  & GRPO & 0.4324 & 0.4686 & 0.7011 & 0.7663 \\
 &  & SRPO & \textbf{0.4330} & \textbf{0.4694} & \textbf{0.7016} & \textbf{0.7676} \\
 \bottomrule
\end{tabular}
}
\end{table}